# Controlling the thermoelectric effect by mechanical manipulation of the electron's quantum phase in atomic junctions

Akira Aiba[1], Firuz Demir[2,3], Satoshi Kaneko[1], Shintaro Fujii[1], Tomoaki Nishino[1], Kazuhito Tsukagoshi[4], Alireza Saffarzadeh[2,5], George Kirczenow[2]*, Manabu Kiguchi[1]*

[1]Department of Chemistry, Graduate School of Science and Engineering, Tokyo Institute of Technology, Ookayama, Meguro-ku, Tokyo 152-8551, Japan

[2]Department of Physics, Simon Fraser University, Burnaby, British Columbia, Canada V5A 1S6

[3]Physics Department, Khalifa University of Science and Technology, P.O. Box 127788, Abu Dhabi, UAE

[4]National Institute for Materials Science, Tsukuba, Ibaraki 305-0044, Japan

[5]Department of Physics, Payame Noor University, P.O. Box 19395-3697 Tehran, Iran

*Correspondence and requests for materials should be addressed to G.K.(E-mail: kirczeno@sfu.ca) and M.K. (kiguti@chem.titech.ac.jp)

**Abstract**

The thermoelectric voltage developed across an atomic metal junction (i.e., a nanostructure in which one or a few atoms connect two metal electrodes) in response to a temperature difference between the electrodes, results from the quantum interference of electrons that pass through the junction multiple times after being scattered by the surrounding defects. Here we report successfully tuning this quantum interference and thus controlling the magnitude and sign of the thermoelectric voltage by applying a mechanical force that deforms the junction. The observed switching of the thermoelectric voltage is reversible and can be cycled many times. Our *ab initio* and semi-empirical calculations elucidate the detailed mechanism by which the quantum interference is tuned. We show that the applied strain alters the quantum phases of electrons passing through the narrowest part of the junction and hence modifies the electronic quantum interference in the device. Tuning the quantum interference causes the energies of electronic transport resonances to shift, which affects the thermoelectric voltage. These experimental and theoretical studies reveal that Au atomic junctions can be made to exhibit both positive and negative thermoelectric voltages on demand, and demonstrate the importance and tunability of the quantum interference effect in the atomic-scale metal nanostructures.

**Introduction**

The quantum nature of fundamental particles is revealed by the wave-like aspects of their behavior. Reflecting the wave-like aspects, quantum interference effects appear in various systems. Quantum interference plays a key role in electron and neutron diffraction by crystals, the Josephson effect in superconductors, electron holography and the Aharonov-Bohm effect in semiconductor nanostructures[1-4]. The applications range from materials science and medicine to quantum computation[5-10]. Quantum effects, including quantum interference, also appear in metal nanostructures[11-17]. When the cross-section of a metal wire bridging two electrodes is reduced to a few atoms, the Fermi wavelength of electrons becomes comparable to the wire's diameter, and quantum mechanical effects govern the electric transport properties of such atomic-scale metal contacts. The conductance of an atomic-scale contact for noble metals such as gold and for alkali metals is quantized in units of $G_0$ ($1G_0 = 2e^2/h = 12.9$ k$\Omega^{-1}$)[11-13]. For such atomic contacts, the quantum interference effect appears as slight fluctuations of the electrical conductance and stronger variations of the thermoelectric voltage[18-22]. The exact values of the electrical properties and their fluctuations depend on the chemical valence of the metal and also on the precise atomic configuration of the atomic-scale metal contact, including the locations of defects, such as grain boundaries that surround the contact. The quantum interference effects of bulk and mesoscopic systems have been utilized[5-8]. However, useful functionalities have not as yet appeared for quantum interference in metal atomic contacts, due to the difficulty in precisely controlling their atomic structures[9,23,24].

Here we demonstrate experimentally that the magnitude and sign of the thermoelectric voltage exhibited by Au atomic junctions can be controlled by varying the mechanical strain applied to the junction. Also, we present numerical quantum transport simulations and a theoretical analysis that show that the quantum interference in metal atomic junctions can be tuned by the application of mechanical strain. We find that the strain tunes the phases of the electronic wave functions in the narrowest part of the junction. This affects the quantum interference throughout the whole device and hence modifies the thermoelectric voltage. The theoretically predicted behavior of the thermoelectric voltage as the strain is varied resembles closely that observed in our experiments. We therefore propose that the physical mechanism that underlies the control of the thermoelectric voltage that we have achieved experimentally is the strain-driven tuning of the electron quantum phases in the narrowest part of the junction that we find in our simulations. Importantly, these effects of varying the strain are demonstrated experimentally and theoretically to be reversible and straining the nanostructure by less than 1 nm results in controlled changes in the thermoelectric voltage, including switching of its sign. By cyclically varying the applied strain, this switching of the thermoelectric voltage can be cycled many times in our experiments. Although polarity changes of the thermoelectric voltage around zero voltage have been reported previously for some Au junctions[18,21,22], no scheme for predictably manipulating the thermoelectric voltage has been reported, due to the difficulties of preparing highly stable metal atomic junctions and fine tuning their atomic configurations. In addition, while the role of quantum interference in the thermoelectric voltages of atomic junctions has been previously studied theoretically[18,22,25], the detailed mechanism through which applied strain modifies the thermopower of Au atomic junctions has not been clear until now. We note that achieving control of the thermoelectric voltage is a necessary prerequisite for any effective thermopower generation using nanoscale devices.

**Results**

In this study, we used the mechanically controllable break junction (MCBJ) approach to study Au atomic junctions, as shown schematically in Fig. 1a (Fig. S1a)[10,17,26]. The Au atomic junctions were prepared by stretching the Au contact by bending the flexible substrate with a piezo element. All experiments were performed in cryogenic vacuum (~20 K). The formation of the Au atomic junction was confirmed by conductance measurements. The last step at 1 $G_0$ ($2e^2/h$) in the conductance trace (Fig. 1b) signals the formation of the Au atomic junction[12,13,17]. Together with the conductance measurement, the thermoelectric voltage $V_T$ of the Au atomic junction was measured by applying a temperature difference across the junction using Pt resistive heaters attached to both ends of the contact (Fig. S2, Fig. S3)[27,28]. The thermoelectric voltage changed simultaneously with the conductance jump from one plateau to the next. In contrast to the monotonic decrease of the conductance, the thermoelectric voltage changed rather irregularly. Even though the conductance stayed nearly constant in the plateau regions, considerable variation in the thermoelectric voltage was observed within the conductance plateaus. Note that the thermoelectric voltage of the Au atomic junction switched between positive and negative values. The large variability of the thermoelectric voltage is evident in the density plot of the thermoelectric voltage against conductance (Fig. 1c). The spread of the thermoelectric voltage values with both positive and negative signs increased with decreasing conductance (and contact size) of the junctions. Although for some Au junctions a polarity change of the thermoelectric voltage around zero voltage has been observed previously, no manipulation scheme of the thermoelectric voltage based on structural and electrical understanding of junction properties has been reported[18,21]. The physical origin of the bipolarity of the thermoelectric voltage of the Au atomic junctions will be discussed below.

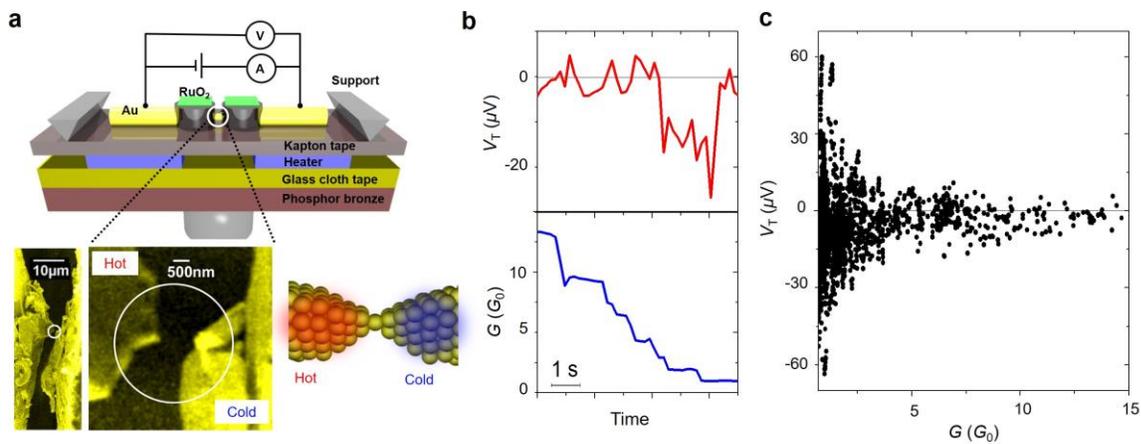

**Figure 1**. **Thermoelectric voltage and simultaneous conductance measurements.** (a) Schematic view of the MCBJ setup used for the simultaneous measurement of the conductance and thermoelectric voltage of the Au atomic junction. The Pt micro heaters and the $RuO_2$ thermometers are placed on each side of the contact: Inset: Scanning electron microscope image of Au contact after it is broken. (b) Conductance and thermoelectric voltage of the Au atomic junction during stretching of the contact. The temperature difference was 16 K and average temperature was 27 K. (c) Distribution of the thermoelectric voltage of the Au atomic junction as a function of conductance of the contact. The distribution is constructed from the combined data of 60 individual curves.

**Mechanical control of the thermoelectric voltage of the Au atomic junction**

We now demonstrate the controllability of the thermoelectric voltage by the mechanical force using the MCBJ device. Figure 2 shows the change in the thermoelectric voltage and conductance together with the stretching/compression distance of the Au atomic junction. Both the conductance and thermoelectric voltage were changed reversibly between two distinct states by varying the mechanical force. It is noteworthy that the polarity of the thermoelectric voltage can be modulated reversibly by the mechanical force. For gold nanostructures it is believed that a conductance $G \sim 2\ G_0$ corresponds to two transport channels (each spin-degenerate) which are realized physically as two Au atoms bridging a junction in parallel, and that $G \sim G_0$ corresponds to one transport channel which is due to a single gold atom bridging the junction[12,13]. Therefore, the switching shown in Fig. 2a corresponds to the transition between a two Au-atom junction and a single gold atom-junction. The switching shown in Fig. 2b corresponds to the transition of a single gold atom junction between two slightly different atomic configurations. In the present switch, the modulation of the thermoelectric voltage was as large as 400 % with respect to the typical thermoelectric voltage of bulk Au, while the conductance changed by (a) 44 % and (b) 5 %. The modulations were calculated using the following equations: $\Delta V_{mod}/\Delta V_{bulk\text{-}Au} = \Delta V_{mod}/(|\text{-}S_{bulk\text{-}Au}\ \Delta T|) = 30/7 = 4.29$ (A) $\Delta V_{mod}/\Delta V_{bulk\text{-}Au}=13/3.1=4.22$, where $S_{bulk\text{-}Au} = 0.7\mu V/K$ and $\Delta T = 10\ K$ or $\Delta T = 4.4\ K$, and $\Delta G_{mod}/\ G = 0.8\ /1.8 = 0.44$ and $0.05/1.0=0.05$.

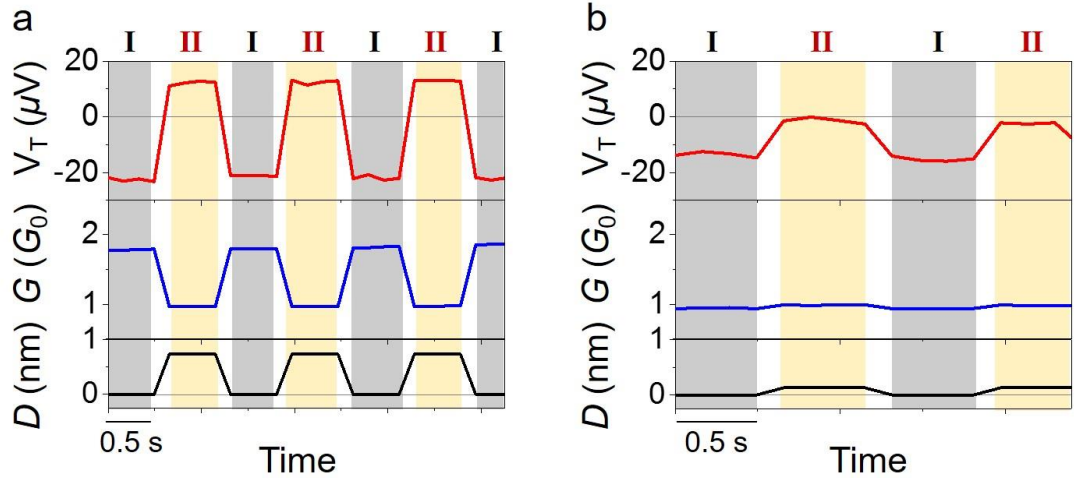

**Figure. 2. Au atomic contact as a thermoelectric voltage switch.** Periodic modulation of the thermoelectric voltage ($V_T$) and conductance ($G$). The modulations were induced by stretching/compressing the junction. The elongation/compression distance ($D$) was (a) 0.73 nm and (b) 0.10 nm. Here, the point of zero elongation ($D$=0) is taken as the most compressed state. The temperature difference was (a) 10 K and (b) 4.4 K, and the average temperature was 25 K.

The reversibility of the observed switching and its reproducibility over multiple cycles seen in Fig. 2 have not been reported in previous experimental studies of metal nanostructures [18,22,23]. Going well beyond previous limited aspects in observation of the polarity change, our observation (together with the theory presented below) strongly suggests the manipulation of the wave function interference in the metallic junction, reproducibly changed by modulation of the junction geometry. Our measurement scheme at low temperatures allowed us to manipulate the stable junction formation. Otherwise the junction could be easily deformed and fluctuating due to thermal activation, resulting in smearing of the quantum effect. Thus we have succeeded in truly controlling the bipolar thermoelectric voltage by an external mechanical force for the first time.

**Quantum interference and the thermoelectric effect**

We propose that the bipolar thermoelectric voltages exhibited by our atomic junctions and their mechanical manipulation that we have demonstrated are the result of modulation of the quantum interference of the wave functions of electrons passing through the junction by the mechanical strains applied to the junction, as will be explained next. At low temperatures, the thermoelectric voltage $\Delta V$ is given by

$$\Delta V = \frac{\pi^2 k_b^2 T}{3e\, \tau(E)} \left. \frac{d\tau(E)}{dE} \right|_{E=E_F} \Delta T \quad (1)$$

where $k_b$, $e$, $E_F$ and $\tau(E)$ are the Boltzmann constant, the electron charge, the Fermi energy and the electron transmission probability at energy $E$ through the nanostructure, respectively[27,29,30]. The thermoelectric voltage is proportional to the energy derivative of the transmission curve at the Fermi level. Consequently its magnitude and sign are sensitive to the energies at which local maxima and minima of $\tau(E)$ are located in the vicinity of the Fermi level.

As was proposed by Ludoph and van Ruitenbeek[18], following previous theoretical work on conductance fluctuations in semiconductor nanoconstrictions[31], these maxima and minima of $\tau(E)$ arise from constructive and destructive quantum interference of the wave functions of electrons passing multiple times through the junction after scattering from remote defects. More recently, there have been numerical quantum transport studies of thermoelectric fluctuations in atomistic models of metal atomic junctions, that also considered the role of defects[22,25]. That random defects can scatter electrons coherently, giving rise to quantum interference phenomena observable in transport measurements on nanoscale and mesoscopic systems at low temperatures is well established both theoretically and experimentally. As well as thermoelectric effects in atomic junctions[18], quantum interference due to coherent scattering by random defects gives rise to such well known phenomena as Anderson localization[32] and universal conductance fluctuations[33]. We note that the presence of defects in the system is important since for an ideal defect-free quantum wire $\tau$ is an integer. Hence for such a system $d\tau/dE = 0$. Therefore the thermoelectric voltage is predicted by Eq. 1 to be zero for an ideal defect-free quantum wire. However, the previous studies of thermoelectric phenomena[18,22,25] did not identify the physical mechanism by which *an applied strain* affects the quantum interference in the device.

Here we identify this physical mechanism: We find that applied strains affect the quantum interference by modifying the phase acquired by an electron passing through the narrowest part of the junction. Depending on the circumstances, the change in the electron's phase induced by stretching the junction can be positive or negative. However, in either case this phase change alters the quantum interference in the device and thus results in a change in the thermoelectric voltage. How this occurs is shown schematically in Fig. 3a and 3b: An electron wave (shown in black) approaches the atomic junction (the green region) from the left and is partly transmitted through the junction (for clarity, waves reflected from the atomic junction are not shown). The transmitted wave is then scattered by a defect (shown in grey) and is partly transmitted to the right (shown in orange) where it exits the system, and partly reflected (shown in red) back through the atomic junction. It is then scattered by another defect (grey) located to the left of the junction and is partly transmitted to the left and partly reflected. The reflected wave (blue) passes again through the junction and exits to the right. In Fig. 3a, the electron Fermi wavelength $\lambda$ and the locations and natures of the defects are such that the transmitted (blue and orange) partial waves that exit to the right are in phase. Thus their interference is constructive and therefore it results in a maximum of the transmission $\tau(E)$ at the Fermi energy, i.e., in a transmission resonance.

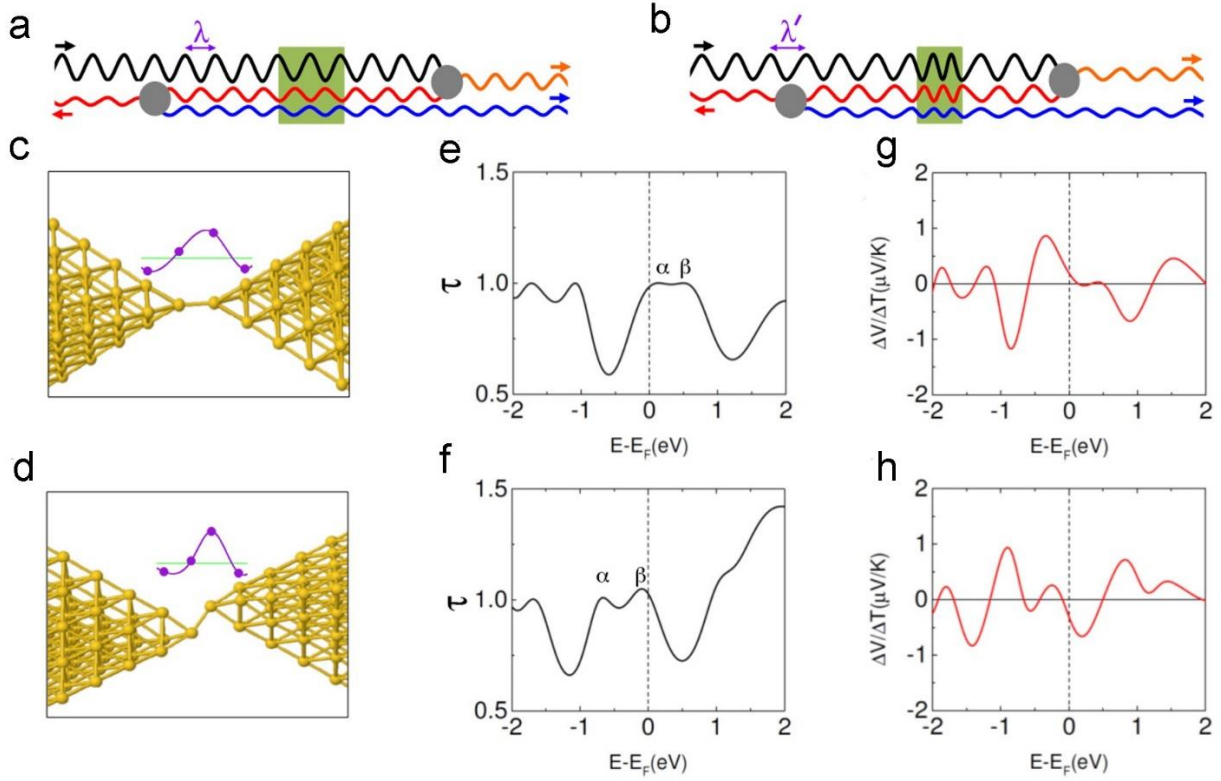

**Figure 3**. **Mechanical manipulation of quantum interference and thermoelectric effect for a weakly strained Au atomic junction.** (a, b): Simplified picture of the quantum interference in Au atomic junctions. An electron wave enters from the left and passes through the atomic junction (shown in green) once, or three times after scattering from defects shown in grey. The wave accumulates different phase shifts $\Delta\varphi = 4\pi$ or $4.5\pi$ as it passes through the unstrained (a) or strained (b) atomic junction, respectively. In both cases the two partial waves exiting to the right interfere constructively (transmission is maximal) because states with suitably differing energies and de Broglie wavelengths ($\lambda$ and $\lambda'$) outside of the junction were chosen. (c, d): Model Au geometries near the junction. In (c) the distance between the two central Au atoms in the junction is 2.75 Å. It has been reduced to 2.63 Å in (d). Insets: Typical calculated wave functions Re($<$6s$|\Psi>$) in the junctions. $\Delta\varphi = 2.23\pi$ and $2.36\pi$ for (c) and (d) respectively. (e, f): Calculated Landauer electron transmission probabilities $\tau(E)$ (ignoring spin) through the structures c, d vs. electron energy $E$. Peaks of $\tau(E)$ are at lower energies in (f) than (e). Conductances are given by $G = G_0\tau(E_F) = 0.985\ G_0$ and $1.03\ G_0$ in (e) and (f), where $G_0 = 2e^2/h$. (g, h): $\Delta V/\Delta T$ defined by Eq. 1 vs. energy $E$ calculated at a mean temperature of 27 K for the structures (c, d). At the Fermi level, $\Delta V = +2.99$ μV and $-4.95$ μV for $\Delta T = 16$ K for (g) and (h).

For slight mechanical modulations of the junction we assume that, in the process of stretching or compressing the junction, the associated structural changes occur mainly at the junction itself, but do not modify appreciably the defects that are located away from the junction. Thus small elastic deformations of the junction are reversible, which explains why the associated changes in thermoelectric voltage are observed to be reversible in our experiments. However, as we shall demonstrate below by numerical simulations, when the junction is deformed, the phase $\Delta\varphi$ accumulated by the electron wave on passing through the junction depends on the applied strain. To illustrate this point and help understand its implications, for definiteness $\Delta\varphi=4\pi$ has been chosen in Fig. 3a, while in Fig. 3b where the atomic junction has been deformed $\Delta\varphi=4.5\pi$ has been chosen. If everything else were to remain the same, this change of $\pi/2$ in $\Delta\varphi$ would result in destructive interference between the two transmitted waves outgoing to the right. However, for gold, lower energy electrons have longer wavelengths. A longer wavelength $\lambda$' outside of the (green) junction region implies a smaller phase shift of the wave as it travels between the junction (green) and the defects (grey) in Fig. 3b. For suitable $\lambda$', this reduced phase shift outside of the atomic junction can compensate exactly for the strain-induced increase in $\Delta\varphi$ in the (green) junction itself and in this way restore the constructive interference between the two partial waves exiting the system to the right. This is the case in Fig. 3b. Thus applying a strain that produces a larger junction phase shift $\Delta\varphi$ implies that constructive interference must occur at a lower energy where the electron de Broglie wavelength in gold is longer. Consequently the peaks of $\tau(E)$ shift to lower energies. This in turn implies a change in the thermoelectric voltage $\Delta V$, and if a peak of $\tau(E)$ shifts from above the Fermi level to below the Fermi level then, according to Eq. (1) the sign of $\Delta V$ is expected to change.

The above physical picture of reversible changes in the thermoelectric voltage $\Delta V$ due to tuning of the quantum interference by mechanical strain can account for our experimental findings. It is supported by our detailed numerical quantum transport calculations (described in detail in the Methods Section) for a *realistic* model of gold atomic junctions. We used density functional theory (DFT) to simulate the structure of the atomic contact and its deformation[34,35]. By solving the Lippmann-Schwinger equation, the electron transmission probability $\tau(E)$ through the junction was calculated as in previous transport calculations for other nanostructures[36-42]. From $\tau(E)$ the conductance and thermoelectric properties of the Au atomic junction were calculated using Eq. (1) and the Landauer formula $G = \tau(E_F)G_0$. The atomic contact is represented by a pair of atomic clusters with up to several hundred gold atoms bridged by a small number gold atoms that form the junction.

We shall now examine theoretically two important cases:

(i) Small strains that do not break bonds between atoms but modify bond lengths and bond angles; experimental data for such a case is shown in Fig. 2b.

(ii) Larger strains involving conformational changes in the junction, such those that result in the larger conductance and thermoelectric voltage changes in Fig. 2a.

**(i) Small strains**

An example of a 240 Au atom structure for which the sign of $\Delta V$ is predicted by our calculations to change when the junction is stretched or compressed by a small amount is shown in Fig. 3c and 3d. The distance between the two central Au atoms in the junction is 2.75 Å in Fig. 3c (the extended structure). In Fig. 3d the electrodes have been moved closer together and displaced laterally; the two central atoms are now separated by 2.63 Å (the compressed structure). The calculated transmission functions $\tau(E)$ for these structures are shown in Fig. 3e and 3f. They exhibit two resonances $\alpha$ and $\beta$ close to the Fermi energy. For the extended structure the Fermi level is located on the lower shoulder of resonance $\alpha$ (Fig. 3e), and $d\tau/dE$ is positive at $E=E_F$ (Fig. 3g). However, for the compressed structure the resonances $\alpha$ and $\beta$ have shifted to lower energies so that the Fermi level is at the upper shoulder of resonance $\beta$ (Fig. 3f), and $d\tau/dE$ is negative (Fig. 3h). Thus, according to Eq. (1), compression of this junction results in reversal of the polarity of thermoelectric voltage $\Delta V$.

That tuning of the quantum interference by the applied strain as sketched out in Fig. 3a and b is the physics responsible for this bipolar behavior of the thermoelectric effect can be seen (Fig. 3c and 3d) by examining the calculated electron wave functions $\Psi$ in the atomic junctions at the Fermi energy. To this end, the real part of the projection of a typical calculated transport wave function $\Psi$ onto the 6s gold atomic orbitals of the atomic junction, $Re(\langle 6s|\Psi\rangle)$, is shown by the purple dots in the insets of Fig. 3c and 3d. (Our justification for choosing the 6s orbital for this purpose is presented in Section 5 of the Supplementary Materials). The purple curves are guides to the eye. Setting <6s|Ψ> = ρexp(iφ), we estimate the wave function phase shifts Δφ when the electron passes through the junction at the Fermi energy to be $2.23\pi$ and $2.36\pi$ for the extended and compressed junctions (Fig. 3c and 3d), respectively. The fact that the phase shift Δφ is larger in Fig. 3d than in Fig. 3c can be seen directly by comparing the insets of these figures; the wavelength in Fig. 3d has clearly been compressed even more than the junction itself has been relative to Fig. 3c, resulting in the additional accumulation of phase in traversing the junction. This difference between Fig. 3c and 3d is qualitatively similar to that between Fig. 3a and 3b. As has been explained above in our discussion of the quantum interference mechanism (Fig. 3a and 3b), the larger phase shift Δφ for the compressed junction is responsible for the shift to lower energies of all of the transmission resonances (including α and β) from Fig. 3e to Fig. 3f. Thus the tunable quantum interference effect is seen to be responsible for the bipolarity of the calculated thermoelectric voltage *ΔV*.

**(ii) Conformation-altering strains**

We have also carried out computer simulations for junctions undergoing more drastic structural changes on elongation and compression and hence exhibiting larger changes in Δφ. We found similar strain-tunable quantum interference to result in bipolar behavior of the thermoelectric voltage in those systems as well, including structures exhibiting thermoelectric voltage and conductance switching consistent with the experimental results presented in Fig. 2a.

Figure 4a shows a model structure where the electrodes are bridged by two Au atoms in parallel, an example of a junction with a conductance not far below $4e^2/h$. In Fig. 4b the electrodes are further apart so that the junction is bridged by two Au atoms in series and its conductance is not far from $2e^2/h$. In Fig. 4c the electrodes are still further apart and the junction is bridged by a chain of three Au atoms so that the conductance is again somewhat below $2e^2/h$. As for the structures in Fig. 3 the geometries of the atoms in the junctions and of nearby atoms have been relaxed to minimize the energy calculated within density functional theory, whereas the atoms further from the junction are assumed to have the structure of bulk fcc gold. Structures broadly similar to those in Fig. 4a-c are believed to have been realized in previous junction stretching experiments[12,13]. We have calculated a series of relaxed intermediate structures (not shown) between that in Fig. 4a and that in Fig. 4b and believe on that basis that, by moving the electrodes appropriately, it is possible to realize the transition between Fig. 4a and b experimentally by straining the junction. We therefore regard the structures in 4a and b as a plausible model for the junctions I and II in Fig. 2a whose experimentally measured conductances are near 2 $G_0$ and $G_0$, respectively.

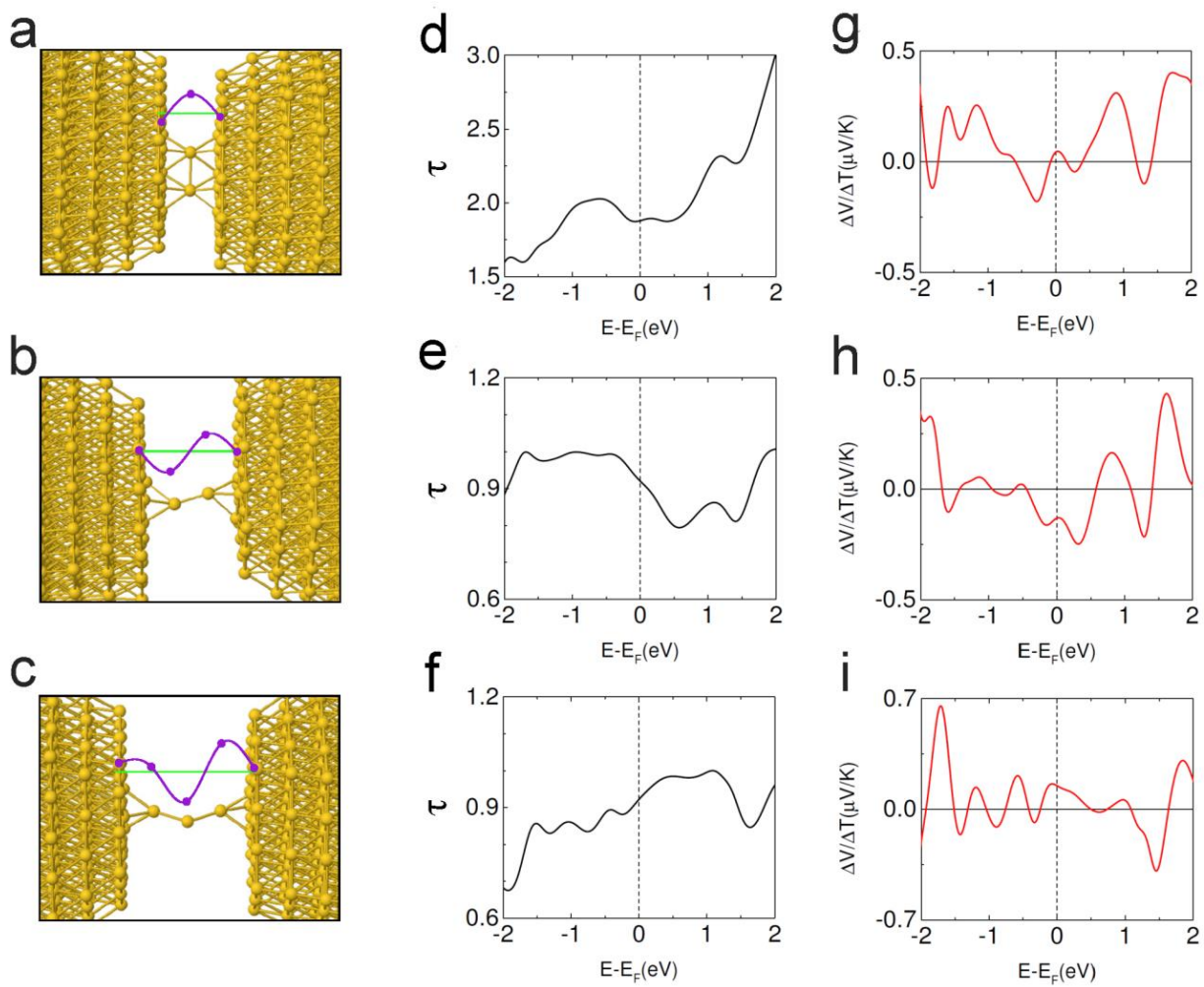

**Figure 4**. **Mechanical manipulation of quantum interference and thermoelectric effect of Au junction undergoing strain-induced conformational changes.** Junctions of two electrodes bridged by (a): two gold atoms in parallel as in Fig. 2b I, (b): two gold atoms in series as in Fig.

2b II, (c): a chain of three gold atoms. The model structures in (a) and (b) each consist of 468 atoms, while structure (c) has 469. Purple dots show calculated values of the real part of a typical electron transport wave function <6s|Ψ> in the junction at the Fermi level. (d-f): The calculated Landauer electron transmission probabilities $\tau(E)$ (ignoring spin) through the structures (a-c) vs. electron energy $E$. The notation is as in Fig. 3. e-f. (g-i): $\Delta V/\Delta T$ defined by Eq. 1 vs. energy $E$ calculated at a mean temperature of 27 K for the structures (a-c). At the Fermi level, $\Delta V = +0.70$ µV, -2.10 µV and +2.40 µV at $\Delta T = 16$ K for (g), (h) and (i) respectively.

The purple dots in the insets of Fig. 4a-c show typical transport wave functions at the Fermi energy in the junctions, represented by Re(<6s|Ψ>) as in Fig. 3. The purple curves are guides to the eye. Here the changes in $\Delta\varphi$ as the structure progresses from Fig. 4a to b to c are several times larger than for the change between the structures in Fig. 3 since here the junction is bridged by increasing numbers of atoms in series whereas in the transition between Fig. 3c and d only bond lengths and bond angles were varied. In Fig. 4a, b and c, $\Delta\varphi = 1.50\pi$, $2.06\pi$ and $2.69\pi$ respectively. These large changes in $\Delta\varphi$ from structure to structure modulate the quantum interference in the device very strongly and consequently result in complete reorganization of the pattern of transmission resonances when the structure changes. Since the resonance landscape around the Fermi level changes drastically from structure to structure, the thermoelectric voltage (including its sign) also changes strongly when such a structural change occurs as can be seen in Fig. 4g-i. In Fig. 4g, h and i the thermoelectric voltage $\Delta V$ is positive, negative and positive respectively for positive $\Delta T$.

We therefore conclude that the strain applied to the atomic junction modulates the quantum interference throughout the device (and hence the thermoelectric voltage) by tuning the phase change that the electron experiences on passing through the narrowest part of the junction both for weak strains that only modify interatomic bond lengths and bond angles and for larger strains that induce conformational changes in the junction.

**Au atomic junction as an electrical conductance and thermoelectric voltage switch**

We evaluated the performance of the thermoelectric voltage switch. The thermoelectric voltage is affected by slight structural changes in the Au atomic junction as demonstrated by the theoretical models (Fig. 3, Fig. 4). Therefore the generated thermoelectric voltage and the magnitude of the mechanical modulations exhibited significant variations between samples. This sample-dependence is consistent with the view that sample-dependent defects play an important role in determining the thermoelectric voltage[18]. However, by breaking and making the junction several times, we obtained Au atomic junctions with desired-performance thermoelectric voltage switching functionality (*i.e.* average thermoelectric voltage and its modulation). Regarding the reproducibility and durability of the thermoelectric voltage switch, the switch can cycle more than 20 times as shown in Fig. 5a. Our results for examples of other samples are presented in Fig. S4. The thermoelectric voltage and conductance switched reversibly between two distinct states. Since the thermoelectric voltage is sensitive to the atomic configuration of the Au atomic junction, this switching behavior indicates that the atomic configuration of the Au atomic junction returned to its initial state after repeated pushing and pulling processes. Transitions between two atomic configurations have been observed previously as a two-level fluctuation of the conductance for Au atomic junctions whose electrode distance was fixed[24]. By contrast, our switch is highly stable. Once the push rod was held in position, the conductance and thermoelectric voltage values were maintained for more than 30 s (Fig. 5b).

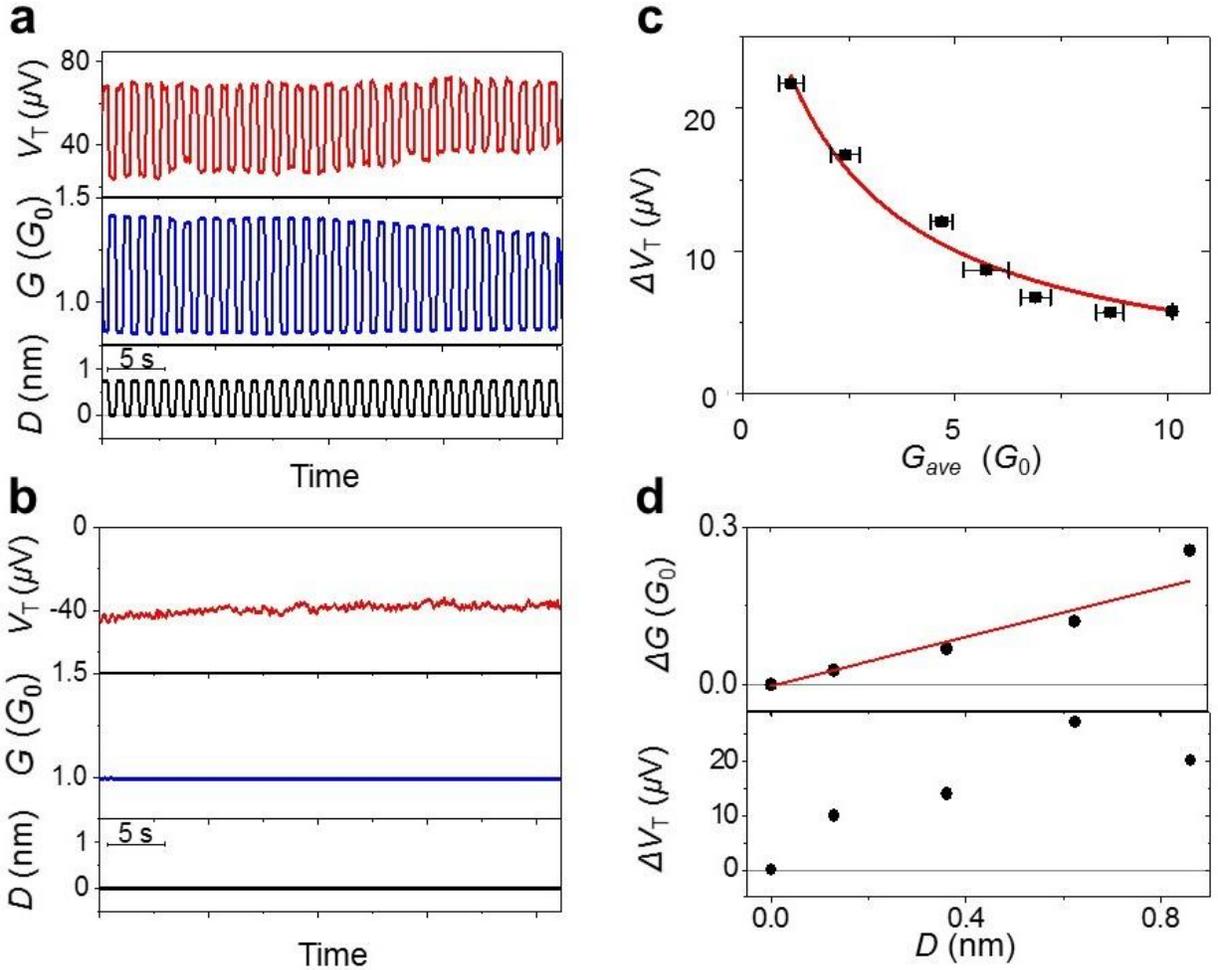

**Figure 5. Performance of the Au atomic junction thermoelectric voltage switch.** (a) Thermoelectric voltage and conductance switching shown together with the displacement of the electrode. (b) Time evolution of the thermoelectric voltage and conductance of the Au atomic contact while the push rod is held in position. The temperature difference is 10 K and average temperature is 40 K. (c) Amount of change in the thermoelectric voltage as a function of conductance of the Au atomic junction. (d) The amount of change in conductance and thermoelectric voltage of the Au atomic junction as a function of the elongation of the contact. The temperature difference was 10 K and average temperature was 25 K.

By breaking and reforming the junction several times, and by changing the separation between the Au electrodes, we obtained Au atomic junctions with switching behavior of the thermoelectric voltage and the electric conductance for displacements under 1 nm. We collected >300 of examples of switching behavior for such displacements, in which the average electric conductance ranged from 1 to 10 $G_0$. The performance of the thermoelectric voltage switch was better in the low conductance regime (see Fig. 5a and Fig. S4). Figure 5c shows a scatter plot of the mechanical modulations of the thermoelectric voltage ($\Delta V_T$) as a function of the conductance ($G$) of the Au atomic junction obtained from >300 examples of the switching behavior. The $\Delta V_T$ increased rapidly with the decrease in the conductance. This is consistent with the fact that the thermoelectric voltage of the Au atomic junction was inversely proportional to the transmission at the Fermi level, as seen in Equation 1. By contrast, the conductance of the Au atomic junction was proportional to the transmission at the Fermi level. Therefore, the thermoelectric voltage decreased with increasing conductance.

Finally, we examine the relationship between the displacement of the junction, and the mechanical modulations of the thermoelectric voltage and the conductance. As shown in Fig. 5d, the amount of change in the conductance increased linearly with increasing displacement. Meanwhile, the change in the thermoelectric voltage was relatively large with respect to the typical thermoelectric voltage of bulk Au and changed nonmonotonically against the displacement. This is a characteristic feature of our thermoelectric voltage switch. As shown in Equation 1, the thermoelectric voltage depends on the slope of the transmission curve at the Fermi level. Compared to the absolute value of the transmission, its slope is more sensitive to structural change of the Au atomic junction. Therefore, a certain change in the thermoelectric voltage was generated even by a slight change in the displacement. As demonstrated by our quantum transport simulation (Fig. 3), an atomic scale electrode displacement causes a small shift in energy of the transmission peaks of the Au atomic junction. Although the conductance is little affected by the small shift, the slope of the transmission peak at the Fermi level is subject to a change of sign.

**Discussion**

We have succeeded for the first time in manipulating the quantum interference of electrons in metal nanostructures, observed as a change in the thermoelectric voltage of the Au atomic junction. Since heating normally destroys quantum interference, it may seem paradoxical at first, that to monitor the quantum interference, we measure voltages that are induced in the Au atomic junction by differential heating of the electrons. However, this thermoelectric voltage measurement is much more sensitive to quantum interference than other experimental methodologies, because the thermoelectric voltage is proportional to the energy derivative of the transmission curve at Fermi level, which is affected by the quantum interference effect as discussed in this paper. The energy derivative is much more sensitive to the change in the transmission curve than the value of the transmission at the Fermi level, that is, the electrical conductance. The higher structural sensitivity of thermoelectric voltage has been also applied to the imaging of the structural disorder in nano materials[43,44]. The thermoelectric measurements amplify the variations in electronic states near Fermi level similarly to a differential filter.

The *reversible* change in the thermoelectric voltage is one of the most interesting new findings in this study. Previously, the appearance of the bipolar thermoelectric voltages in the form of fluctuations of the thermoelectric voltage have been reported for metal atomic contacts, and conductance switching has been reported for the Au atomic contacts[18,22,23]. But, a successful thermoelectric voltage switch has never been reported for a metal atomic contact, for the following reason. The thermoelectric voltage is sensitive to the distribution of defects located away from the junction in the case of the Au atomic junction. The reversible change requires that the atomic configuration including the distribution of defects returns to its initial state when mechanical strain is removed. In our experiments, the small mechanical modulation of the Au atomic junction and low temperature measurement enable the realization of this seemingly unlikely scenario. The associated structural changes occur only at the junction itself but do not modify the defects that are located away from the junction because the stresses (and hence also the strains) induced in the structure by the applied external force are largest where the junction is narrowest. For these reasons the deformation of the atomic junction and of the device as a whole is fully reversible. This property of the elastic behavior of atomic junctions is expected to be quite general and to hold for junctions formed from many different materials. Since it is key to our ability to manipulate quantum interference and hence the thermoelectric effect in our atomic junctions, it is reasonable to expect similar manipulation to be achievable in atomic junctions formed from other substances. From a practical perspective, achieving control of the thermoelectric voltage is a necessary prerequisite for any effective thermopower generation using nanoscale devices.

We have successfully manipulated the quantum interference of electrons in metal nanostructures. The quantum interference is detected as a change in the thermoelectric voltage and electrical conductance of the Au atomic junction. By the application of a mechanical force that elongates or compresses the junction we are able to tune the phases of the wave functions of electrons passing through the junction and thus tune the electron quantum interference in the metal nanostructures as a whole. This in turn tunes the thermoelectric voltage and can change its sign. This process is reversible, i.e., by reversing the force, the electronic properties are returned to their original values. Our *ab initio* and semi-empirical calculations of the structure and quantum transport in atomic junctions confirm that quantum interference in atomic junctions can be tuned reversibly by deforming the junction and that this results in reversal of the thermoelectric voltage for realistic deformations of the atomic junction. The experimental manipulation of quantum interference and thermoelectricity that we have achieved in Au atomic junctions depends on a general property of the spatial distribution of strains in deformed atomic-scale junctions. It is therefore reasonable to expect similar control to also be achievable for atomic junctions formed from other materials.

**Methods**

**Experimental method.**   The measurements were performed using the mechanically controllable break junction (MCBJ) technique (Fig. 1a). The MCBJ substrate was fabricated by the following process. Two Pt resistive heaters (width 200 μm, length 8 mm, thickness 120 nm) were deposited on polyimide tape via magnetron sputtering. The tops of the heaters were covered with polyimide tape for electrical insulation. The covered Pt heaters and the phosphor bronze substrate were thermally insulated with fluoroplastic saturated glass cloth tape. A notched Au wire (0.1 mm diameter, 99.99% purity) was fixed with epoxy adhesive (Stycast 2850FT) on the covered Pt heaters. The RuO$_2$ thermometers were fixed on the Au wire at both ends (Fig. S1). The MCBJ sample was mounted in a regular MCBJ setup inside a vacuum chamber.

The electric measurements were performed with a source measure unit (Keithley 2612A). The conductance and thermoelectric voltage were measured in four steps at low temperatures (~50 K), as follows: 1) The voltage was measured at zero current bias. 2) The current was measured at a voltage of +50 mV. 3) The voltage was measured at zero current bias. 4) The current was measured at a voltage of -50 mV. Each cycle took approximately 0.2 s. The conductance was obtained from the current difference for the two voltage polarities, and the thermoelectric voltage was obtained from the average of two voltage values at zero bias current.

**Theoretical method.** In the models that we consider the electrodes are represented by a pair of atomic clusters with up to several hundred gold atoms bridged by a small number gold atoms that form the junction. The positions of the gold atoms in the junction and in its immediate vicinity were estimated by minimizing the total energy of the system, computed within density functional theory using the GAUSSIAN 09 package with the PBE1PBE functional and Lanl2DZ pseudopotentials and basis sets[34,35]. The atoms of the electrodes further from the junction were fixed during the relaxations and assumed to have the structure of fcc bulk gold.

In order to compute $\tau(E)$ for these model structures, we attached a large number of semi-infinite quasi-one-dimensional ideal leads representing electron reservoirs to the valence orbitals of the outer atoms of the gold clusters that represent the two electrodes, as in previous work on transport through nanostructures[36-42]. In the present model, the boundaries between these ideal leads and the gold clusters may be viewed as playing the role of the electrode grain boundaries that are closest to the atomic junction in the experimental system. We note that in our experimental samples other defects such as vacancies, staking faults and interstitials may also be expected to be present, in addition to grain boundaries. The electron transmission probability $\tau(E)$ through the system consisting of the electrodes and junction was obtained by solving the Lippmann-Schwinger equation

$$\left|\Psi^\alpha\right\rangle = \left|\Phi_0^\alpha\right\rangle + G_0(E) W \left|\Psi^\alpha\right\rangle \quad (2)$$

for the scattering states of electrons traveling through the junction. Here $\left|\Phi_0^\alpha\right\rangle$ is an electron eigenstate of the $\alpha^{th}$ ideal semi-infinite one-dimensional lead that is decoupled from the gold

clusters representing the electrodes, $G_0(E)$ is the Green's function of the decoupled system, $W$ is the coupling between the electrodes and the ideal leads, and $|\Psi^\alpha\rangle$ is the scattering eigenstate of the complete coupled system associated with the incident electron state $|\Phi_0^\alpha\rangle$. The semi-empirical extended Hückel model[41] with the parameters of Ammeter *et al.*[45,46] was used to evaluate the Hamiltonian matrix elements and atomic valence orbital overlaps that enter the Green's function $G_0(E)$ in Eq. (2). As has been discussed in Refs. 37 and 41, this methodology involves no fitting to any experimental data relating to transport. As well as yielding the conductance quantization observed in gold atomic wires, it is known to yield low bias conductances in reasonably good agreement with experiments for a variety of molecular wires bridging gold electrodes[36,37,41,42,47-50].

**Acknowledgements**

This work was supported by JSPS KAKENHI Grant Numbers 26102013, 16K13975, Asahi glass, Murata, Mitsubishi and Iketani foundation, NSERC, CIFAR, Westgrid, and Compute Canada.

**Author Contributions**

The experiments were conceived and carried out by A.A., S.K., S.F., T.N., and M.K. The calculations were carried out and their physical interpretation was provided by F.D, A.S and G.K. The manuscript was written by A.A, S.K. K.T. M.K. and G.K., with comments and input from all authors